\documentclass[english,useAMS,usenatbib]{mn2e}
\usepackage{fancyhdr}
\usepackage{float}
\usepackage{subfig}
\usepackage[T1]{fontenc}
\usepackage[latin9]{inputenc}
\usepackage{units}
\usepackage{rotating}
\usepackage{url}
\usepackage{amsmath}
\usepackage{amssymb}
\usepackage{graphicx}
\usepackage{esint}
\usepackage{hyperref}
\usepackage[para,online,flushleft]{threeparttable}

\makeatletter

\def\gtsima{$\; \buildrel > \over \sim \;$}
\def\ltsima{$\; \buildrel < \over \sim \;$}
\def\prosima{$\; \buildrel \propto \over \sim \;$}
\def\gsim{\lower.5ex\hbox{\gtsima}}
\def\lsim{\lower.5ex\hbox{\ltsima}}
\def\simgt{\lower.5ex\hbox{\gtsima}}
\def\simlt{\lower.5ex\hbox{\ltsima}}
\def\simpr{\lower.5ex\hbox{\prosima}}
\def\la{\lsim}
\def\ga{\gsim}

\providecommand{\tabularnewline}{\\}

\title[Cosmic archaeology with stellar BH binaries]{Cosmic archaeology with massive stellar black hole binaries}
\author[L. Graziani et al.]{L. Graziani$^{1,2,3}$
\thanks{E-mail: luca.graziani@roma1.infn.it}, R. Schneider$^{1,2,4}$, S. Marassi$^{1,2}$, W. Del Pozzo$^{5}$, M. Mapelli$^{6,7,8}$, 
\newauthor N. Giacobbo$^{6,7,8}$\\
$^{1}$Dipartimento di Fisica, Sapienza, Universit$\grave{a}$ di Roma, Piazzale Aldo Moro 5, 00185, Roma, Italy\\
$^{2}$INFN, Sezione di Roma I, P.le Aldo Moro 2, 00185 Roma, Italy\\
$^{3}$INAF/Osservatorio Astrofisico di Arcetri, Largo E. Femi 5, 50125 Firenze, Italy \\
$^{4}$INAF/Osservatorio Astronomico di Roma, Via di Frascati 33, 00078 Monte Porzio Catone, Italy\\
$^{5}$Dipartimento di Fisica ``Enrico Fermi'', Universit\`a di Pisa, Pisa I-56127, Italy \\
$^{6}$Dipartimento di Fisica e Astronomia ''G. Galilei'', Universit\`a di Padova, vicolo dell'Osservatorio 3, 35122 PD, Italy\\
$^{7}$INFN, Sezione di Padova, Via Marzolo 8, I-35131 Padova, Italy \\
$^{8}$INAF/Osservatorio Astronomico di Padova, Vicolo dell' Osservatorio 5, I-35122, Padova, Italy}

\usepackage{babel}
\usepackage{natbib}

\begin{document}

\date{Accepted XX <Month> XX. Received 2019 <Month> XX; in original form
2019 <Month> XX}

\maketitle
\pagerange{\pageref{firstpage}--\pageref{lastpage}} \pubyear{2020}\label{dat:firstpage}

\begin{abstract}

The existence of massive stellar black hole binaries (MBHBs), with primary black hole (BH) masses $\ge 31 \, M_\odot$, was proven by the detection of the gravitational wave (GW) event GW150914 during the first LIGO/Virgo observing run (O1), and successively confirmed by seven additional GW signals discovered in the O1 and O2 data. By adopting the galaxy formation model \texttt{GAMESH} coupled with binary population synthesis (BPS) calculations, here we investigate the origin of these MBHBs by selecting simulated binaries compatible in mass and coalescence redshifts. We find that their cosmic birth rates  peak in the redshift range $6.5 \leq z \leq 10$, regardless of the adopted BPS. These MBHBs are then old systems forming  in low-metallicity ($Z \sim [0.01-0.1] \, Z_{\odot}$), low-stellar-mass galaxies, before the end of cosmic reionization, i.e. significantly beyond the peak of cosmic star formation. GW signals generated by coalescing MBHBs open up new possibilities to probe the nature of stellar populations in remote galaxies, at present too faint to be detected by available electromagnetic facilities.
\end{abstract}

\begin{keywords} Cosmology: theory - Cosmic UV background - IGM - metal ions \end{keywords}

\section{INTRODUCTION\label{sec:INTRODUCTION}}

Since the discovery of the first GW signal GW150914 \citep{2016PhRvL.116f1102A} and to date, the LIGO/Virgo 
Collaboration detected four events interpreted as originated by the coalescence of MBHBs (i.e. systems with m$_{1} \in [31,66]$~M$_{\odot}$, m$_{2} \in [21,43]$~M$_{\odot}$) at a median luminosity distance $\rm d_L \ge 440$~Gpc \citep{2019PhRvX...9c1040A}; interestingly enough, a recent independent data analysis \citep{2019arXiv190407214V} expanded the above sample with four new systems (see Table \ref{tab:BBH_obs}). Even more intriguing, the current O3 run has already reported more than fourteen alerts with similarly high $\rm d_L$\footnote{https://gracedb.ligo.org/latest/}.

\begin{table*}     
\centering          
\begin{tabular}{c|c|c|c|c|c|c|c|c}
\hline\hline
    & GW150914 & GW170729 & GW170818 & GW170823 & {\bf GW170304} & {\bf GW170403} & {\bf GW170425} & {\bf GW170727}\\ 
\hline
$\rm m{_1}/M_{\odot}$ & $35.6^{+4.8}_{-3.0}$ & $50.2^{+16.2}_{-10.2}$ & $35.4^{+7.5}_{-4.7}$ & $39.5^{+11.2}_{-6.7}$ & $41.0^{+12.0}_{-7.0}$ & $44.0^{+12.0}_{-8.0}$ & $44.0^{+19.0}_{-10.0}$ & $39.0^{+10.0}_{-6.0}$ \\
\hline
$\rm m{_2}/M_{\odot}$ & $30.6^{+3.0}_{-4.4}$ & $34.0^{+9.1}_{-10.1}$  & $26.7^{+4.3}_{-5.2}$ & $29.4^{+6.7}_{-7.8}$ & $31.0^{+7.0}_{-8.0}$ & $32.0^{+8.0}_{-9.0}$ & $29.0^{+11.0}_{-8.0}$ & $29.0^{+6.0}_{-7.0}$ \\
\hline
$\rm \mathcal M/M_{\odot}$ & $28.6^{+1.7}_{-1.5}$ & $35.4^{+6.5}_{-4.8}$ & $26.5^{+2.1}_{-1.7}$& $29.2^{+4.6}_{-3.6}$ & $47.0^{+8.0}_{-7.0}$ & $48.0^{+9.0}_{-7.0}$ & $47.0^{+26.0}_{-10.0}$ & $42.0^{+6.0}_{-6.0}$\\
\hline
$\rm M_{f}/M_{\odot}$ & $63.1^{+3.4}_{-3.0}$ & $79.5^{+14.7}_{-10.2}$ & $59.4^{+4.9}_{-3.8}$ &$65.4^{+10.1}_{-7.4}$ & $^{}_{}$ & $^{}_{}$ & $^{}_{}$ & $^{}_{}$\\
\hline
$\rm d_{L}/Mpc$ & $440^{+150}_{-170}$ & $2840^{+1400}_{-1360}$ & $1060^{+420}_{-380}$ & $1940^{+970}_{-900}$ & $^{}_{}$ & $^{}_{}$ & $^{}_{}$ & $^{}_{}$ \\
\hline
$z_c$           & $0.09^{+0.03}_{-0.03}$ & $0.49^{+0.19}_{-0.21}$ & $0.21^{+0.07}_{-0.07}$ & $0.35^{+0.15}_{-0.15}$ & $0.50^{+0.2}_{-0.2}$ & $0.45^{+0.22}_{-0.19}$ & $0.50^{+0.4}_{-0.3}$ & $0.43^{+0.17}_{-0.17}$ \\ 
\hline
${\cal R}$ (SeBa/MOBSE) & 2.01/87.26 & 0.06/6.81 & 5.23/124.90 & 5.65/111.10 & 7.14/52.34 & 6.99/49.12 & 7.38/108.63 & 6.37/112.39\\ 
\hline\hline
\end{tabular}
\\\vspace{2mm}
\caption{Properties of the GW events (in column) associated with the MBHBs found in O1 and O2 \citep{2019PhRvX...9c1040A} and extended by \citet{2019arXiv190407214V} (although with lower $p_{\rm astro}$ and FAR values, GW IDs in bold). Each row  shows: source frame component masses $\rm m_{1}$ and $\rm m_{2}$, chirp mass $\mathcal {M}$, final source frame mass $\rm M_{f}$, luminosity distance $\rm d_{L}$, estimated coalescence redshift $z_c$ and estimated coalescence rates (${\cal R}$~[cGpc$^{-3}$ yr$^{-1}$]) predicted by \texttt{SeBa}/ \texttt{MOBSE} in our Local Group-like volume of $4^3$~cMpc$^3$.} 
\label{tab:BBH_obs}
\end{table*}

Future ground-based interferometers, such as KAGRA \citep{Akutsu:2018axf} and LIGO-India will join the global GW detector network improving the event localization up to $90\%$-confidence \citep{Abbott2018}. Space-based missions will target the milli-Hz band with LISA\footnote{https://www.elisascience.org/} and the deci-Hz band with DECIGO\footnote{http://tamago.mtk.nao.ac.jp/decigo/}. This synergistic multi-band approach \citep{2016PhRvL.116w1102S} will place better constraints on MBHBs, also accessing their early inspiral phases \citep{2018PTEP.2018g3E01I}. Finally, 3G detectors, such as the Einstein Telescope\footnote{http://www.et-gw.eu/} and Cosmic Explorer\footnote{https://cosmicexplorer.org/} could detect stellar MBHBs up to extremely high redshifts \citep{kalogera2019}.

Stellar models predict MBHBs to be the end products of metal-poor stars \citep{MAPELLI2009, MAPELLI2010, Belczynski2010, Spera2015}. Given our current understanding of galaxy evolution, these stars are preferentially formed in low-mass and less-chemically-evolved galaxies \citep{Maiolino2019}, hardly resolved by large scale cosmological simulations.

BPS codes are traditionally adopted to investigate the evolution of BH binaries by generating databases (DB) from distributions of initial stellar masses and orbital parameters. By coupling them with estimates of the cosmic star formation rate (SFR) and of the average metallicity evolution (or mass-metallicity relation), their coalescence rates along $z$ can be inferred  
\citep{Schneider2010, Marassi2009, MARASSI2011, REGIMBAU2011, DOMINIK2013, Belczynski2016, Lamberts2016, Dvorkin2016, Elbert2018, Chruslinska2019, Neijssel2019, Bavera2019}. With hydrodynamic simulations or semi-analytic models (SAMs) the cosmological evolution of compact binaries can be studied connecting galaxies hosting their birth and coalescence \citep{Schneider2017,MAPELLI2017, OShaughnessy2017, Mapelli2018, Marassi2019,Artale2019}. 

Here we use the \texttt{GAMESH} model to predict the origin of MBHBs in a Local Group-like volume. In \citet{Schneider2017} we already investigated the birth and coalescence sites of compact binaries generating O1 GW events, while in  \citet{Marassi2019} we looked at observational counterparts of GW150914 hosts. Here, we go one step forward by exploring the birth and coalescence of the MBHBs in Table \ref{tab:BBH_obs}, with an increased statistical sample of massive binaries and by comparing predictions of two independent BPS databases (DB): \texttt{SeBa} \citep{PortegiesZwart1996, MAPELLI2013} and \texttt{MOBSE} \citep{Giacobbo2018}.

We provide the statistical evidence that the highest birth rate of their stellar progenitors is found in low metallicity ($Z \leq 0.1$~Z$_{\odot}$), star forming, dwarf galaxies living in the redshift range $6.5 \leq z\leq 10$, i.e. in the epoch of reionization (EoR, $z \geq 6$). While this result is proven to be independent of the adopted BPS, the number of coalescence events strongly depends on the prescriptions implemented in binary evolution codes for massive BH formation.

\section{Galaxy formation model}
\label{sec:Model}
\texttt{GAMESH} \citep{Graziani2015, Graziani2017, Graziani2019} is a galaxy formation model based on a hybrid pipeline combining a Dark Matter (DM)-only simulation, a SAM for star formation and chemical evolution and a  radiative transfer module. The DM run simulates a multi-zoom cosmic box better resolved in its inner cubic volume of $4^3$ cubic comoving mega-parsecs (cMpc$^3$), centered on a Milky Way-like halo (a Local Group-like volume). The SAM module runs on a galaxy catalog taken from a larger volume ($\sim 8^3 \,$~cMpc$^3$) to capture a wider statistics of intermediate/dwarf galaxies whose stellar and chemical evolution in $0 < z <20$ is regulated by two parameters: star formation and wind efficiency. The resulting baryonic properties of the MW are in agreement with observations. Moreover, the histories of a plethora of well resolved dwarf galaxies, co-evolving under strong dynamical interactions and feedback\footnote{Here  radiative feedback is implemented by assuming that the volume instantly re-ionizes at $z=6$.}, naturally reproduce observed galaxy scaling relations  \citep{Graziani2017,Ginolfi2018}.  
In \citet{Schneider2017} \texttt{GAMESH} was extended to self-consistently account for compact binary systems by assigning a binary fraction of $f_{2,*}=1$ and by randomly sampling the newly formed stellar mass with a \texttt{SeBa} DB having  $2\times 10^6$ binaries in the IMF mass range M$_{\star} \in [0.01, 100]$~M$_{\odot}$. Here we adopt two new  independent DBs improving the statistics of MBHBs: a \texttt{MOBSE} DB with $10^7$ binaries sampling M$_{\star} \in [5.0,150]$ and a \texttt{SeBa} DB with $2 \times 10^7$ systems in M$_{\star} \in [8.0,100]$ . Each DB has 12 metallicity bins, regularly spanning the range $Z\in[0.01, 1]$~Z$_{\odot}$\footnote{For details on the set-up of the two DBs, please refer to the $\alpha$5 run of \citet{Giacobbo2018b} and to \citet{Schneider2017}. We also assume $Z_\odot = 0.02$. Finally note that the prescriptions in MOBSE and SeBa are not tailored to describe systems to describe systems with $Z < 0.01$~Z$_{\odot}$ and we are forced to extrapolate the results of $Z = 0.01 Z_\odot$ at lower metallicities, with a possible impact on the results, especially for galaxies hosting Pop\ III stars.}
While the two BPS assume the same stellar evolutionary tracks and metallicity dependent mass loss in stellar winds\footnote{In \texttt{MOBSE} the metallicity dependence is also suppressed when the electron scattering Eddington factor $\Gamma_{\rm e} \ge 2/3$ \citep{Giacobbo2018}.}, the stellar evolution channels producing massive BHs are significantly different: the \texttt{MOBSE} $\alpha$5 run adopts the rapid SN model of \citet{Fryer2012}, while in \texttt{SeBa} all stars with pre-SN masses $m_{\rm pre,SN} \ge 40 M_\odot$ are assumed to collapse into a BH with no SN explosion; the resulting BH mass is then $m_{\rm BH} = m_{\rm CO} + (2/3) (m_{\rm He} + m_{\rm H})$ \citep{MAPELLI2013}. The Common Envelope (CE) efficiencies also differ: $\alpha = 1.0$, $\lambda = 0.5$ in \texttt{SeBa}, while in \texttt{MOBSE} $\alpha = 5$ and $\lambda$ depends on the stellar type. Note that these parameters critically affect the statistics of low-mass BHBs but have a minor impact on the merger rate of MBHBs \citep{Giacobbo2018}.

\begin{figure}
\centering
\vspace{\baselineskip}
\includegraphics[angle=0,width=0.60\textwidth,trim={1.2cm 0cm 0cm 0.25cm},clip]{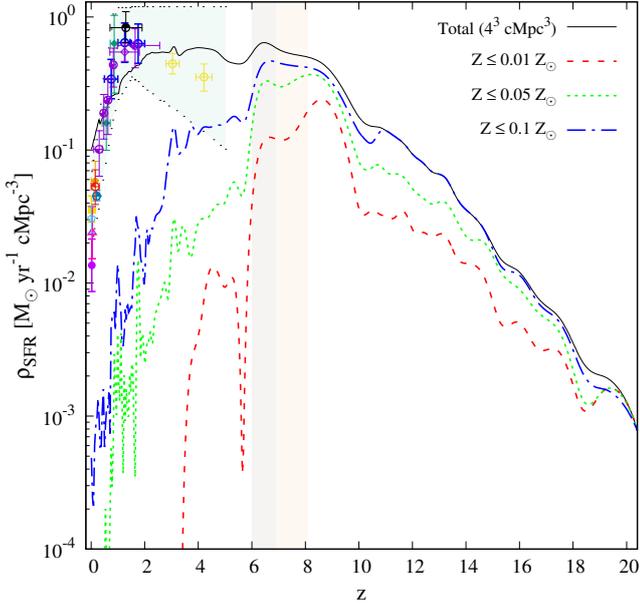}
\caption{Star formation rate density as a function of redshift $z$ in the ($4$~cMpc$)^3$ volume: total (solid black), from galaxies with $Z_{\star}\leq 0.01 Z_{\odot}$ (dashed red line), $Z_{\star}\leq 0.05$~$Z_{\odot}$ (dotted green), and $Z_{\star}\leq 0.1 Z_{\odot}$ in dotted-dashed blue line. Shaded areas indicate the redshift range of the reionization midpoint $6.9 < z_{50\%} < 8.1$ (light pink, \citet{Planck2018}) and the assumed end of reionization (gray). Observational data and its general level of uncertainty in the Local Volume (cyan shaded area) are collected from \citet{Hopkins2001}.}
\label{fig:Fig1}
\end{figure}

Once the newly formed M$_{\star}$ in each galaxy is populated with binaries randomly sampled from the DB with $Z$ closest to the stellar metallicity $Z_{\star}$, we follow them in time from their birth ($t_0$) to coalescence ($t_c$), by relating ancestors with descendant galaxies. 
In Fig. \ref{fig:Fig1} we show the predicted total SFR density $\rho_{\rm SFR}$ as a function of redshift (solid black line). Dashed-red, dotted-green and dashed-dotted blue lines correspond to the same quantity computed by summing up the contributions of star forming galaxies with $Z_{\star} \leq 0.01, 0.05, 0.1$~$Z_{\odot}$, respectively. The simulated trend is in very good agreement with the observational data at $z < 4$ and their range of uncertainty, collected from \citet{Hopkins2001}. Noticeably, systems with $Z \leq 0.05$~$Z_{\odot}$ provide a major contribution to $\rho_{\rm SFR}$ from the cosmic dawn ($z\sim 18$) down to $z\sim 6$, making small, normal star forming galaxies (see \citealt{2020MNRAS.tmp..726G} for a definition) the dominant population at these epochs. Gas photo-heating associated to cosmic reionization progressively diminishes their contribution (see the relative drops in red/green/blue lines) until the total SFR becomes sustained only by intermediate-mass galaxies hosted in Ly$\alpha$-cooling halos at $z<6$ \citep{Graziani2015}. 
 
\section{Results}
\label{sec:Results}

Before presenting our results, we note here that MBHBs are identified in the simulation by requiring that both masses, $m_1,m_2$, and coalescence redshift $z_c$ (derived from $t_c$) lie within the observational uncertainties reported in Table \ref{tab:BBH_obs}. 

\subsection{MBHBs formation sites and birth rates}
\label{sub:birth}

\begin{figure}
\centering
\vspace{\baselineskip}
\includegraphics[angle=0,width=0.49\textwidth,trim={0.15cm 0cm 0 0.16cm},clip]{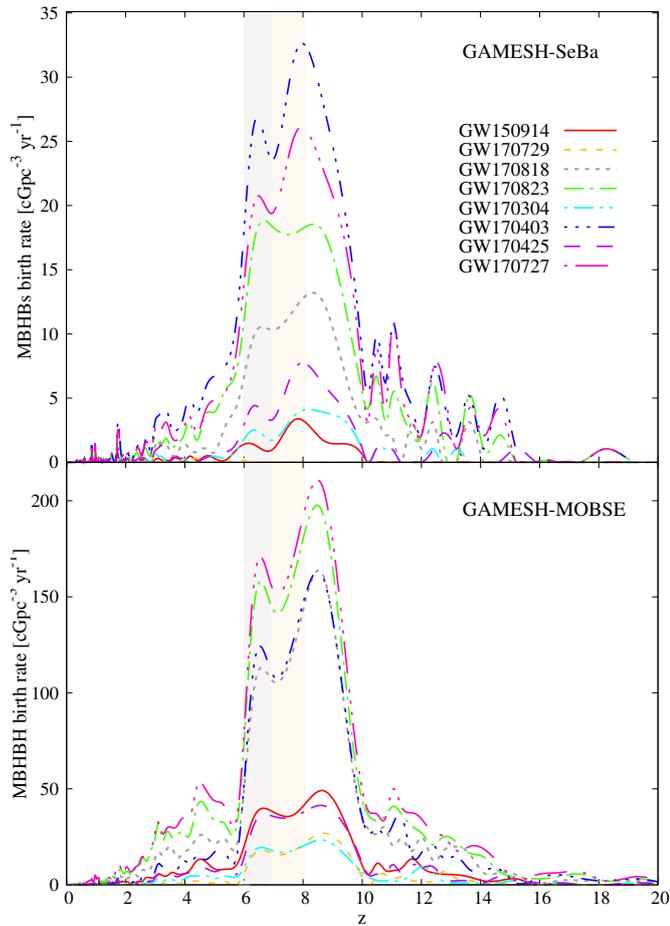}
\caption{Birth rates [cGpc$^{-3}$ yr$^{-1}$] of our MBHBs stellar progenitors as a function of redshift $z$. The two panels adopt the same galaxy formation model but different BPS calculations: \texttt{SeBa} (top) and \texttt{MOBSE} (bottom). The light pink and gray shaded areas are the same as in Fig. \ref{fig:Fig1}.}
\label{fig:Fig2}
\end{figure}

The birth rates of stellar progenitors evolving into the selected MBHBs are shown in Fig. \ref{fig:Fig2}, as a function of $z$; top/bottom panels show rates obtained coupling with \texttt{SeBa}/ \texttt{MOBSE} with identical line styles and colours for the same GW signal. It is immediately evident that all birth rates peak in the redshift range $6.5 \leq z \leq 10$ regardless the adopted BPS, and that their shape is similar across GW signals, reflecting the underlying SFR($z$) trend\footnote{This result is peculiar of the selected massive binaries. A broader mass selection extending to lower BH masses would shift their birth rates closer to the SFR peak. Moreover, our previous results \citep{Schneider2017}, while based on the same cosmological run, did not have enough statistical sampling of the high-mass end of the stellar initial mass function and therefore underestimated the birth rates of GW150914-like events in low-mass, low-metallicity galaxies at high $z$ (see Section \ref{sec:Model}). Finally, the results are also confirmed by Monte Carlo convergence tests performed with different random number chains.}. 
The absolute values for each signal, on the other hand, strongly vary across BPS predictions as well as their relative height and line shapes (see also Section \ref{sub:Z}). 
The coalescence rate of each MBHB is provided in the last row of Table \ref{tab:BBH_obs}, while the total merger rates (i.e. when all binary BHs in the simulation are considered, regardless of their masses) at $z=0.2$ and $z=0$ are ${\cal R}_0 = 4195$ (1513) Gpc$^{-3}$ yr$^{-1}$ and ${\cal R}_{0.2} = 5564$ (1584) Gpc$^{-3}$ yr$^{-1}$ for \texttt{SeBa} (\texttt{MOBSE}); consequently our MBHBs contribute only for 7 (41)\% to the total value at $z=0$. 

While a direct comparison with observationally inferred rates [$24.4 \-- 140.4$]~Gpc$^{-3}$ yr$^{-1}$ \citep{Abbott2019} is not feasible because Local Group-like volumes are generally over-dense and then not representative of larger cosmological scales\footnote{The $\rho_{\rm SFR}$ shown in Fig \ref{fig:Fig1} is approximately one order of magnitude larger than the cosmic star formation rate density at $z < 4$ \citep{Madau2017} and flatter at higher $z$.}, we note that \citet{MAPELLI2017} adopted the same \texttt{MOBSE} DB on the Illustris simulation (with a cubic box size of $L_{\rm box}$ = 106.5 cMpc) finding ${\cal R}_0 = 155$ Gpc$^{-3}$ yr$^{-1}$ and ${\cal R}_{0.2} = 228$ Gpc$^{-3}$ yr$^{-1}$, close to the 90\% credible values of \citet{Abbott2019}.
However, the contribution of high-$z$ dwarfs remains mostly undetermined in large cosmological simulations and in models that adopt observationally inferred scaling
relations, such as the mass-metallicity relation and galaxy main sequence, which are not yet observed at $z \geq 6$.

\begin{figure*}
\centering
\vspace{\baselineskip}
\includegraphics[angle=0,width=0.88\textwidth,trim={0 0 0 0.4cm},clip]{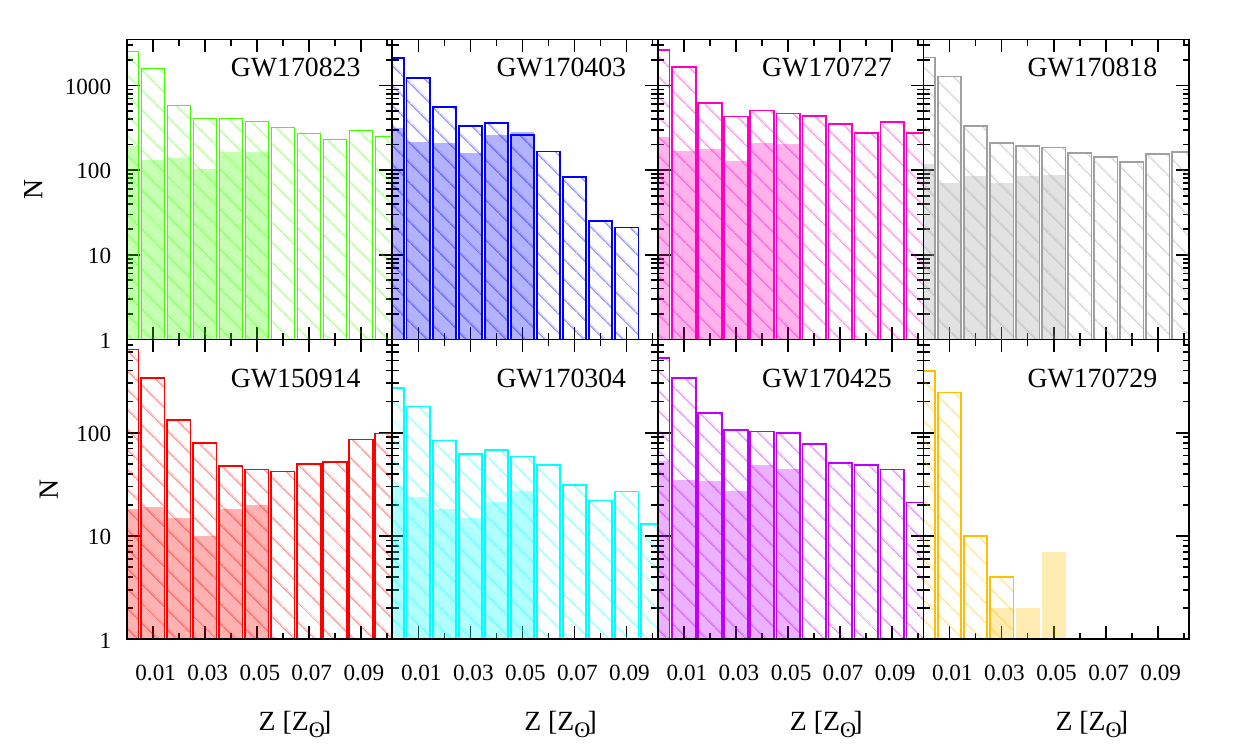}
\caption{Total number of GW events (N, in logarithmic scale) predicted to form at each $Z$ by \texttt{SeBa} (filled histograms) and \texttt{MOBSE} (dashed histograms) run on the same galaxy formation model.
Each panel represents the results obtained for individual GW events and the colour coding is the same in Fig. \ref{fig:Fig2}.}
\label{fig:Fig3}
\end{figure*}

\subsection{Metallicity dependence}
\label{sub:Z}

The distribution in metallicity of MBHBs stellar progenitors is shown in Fig. \ref{fig:Fig3}; filled (dashed) histograms show the results obtained with \texttt{SeBa} (\texttt{MOBSE}). All the stellar progenitors predicted with the \texttt{SeBa} DB form at metallicity $Z \leq 0.05 Z_\odot$ following a nearly flat distribution\footnote{For GW170729, the most massive among the MBHB sample shown in Table \ref{tab:BBH_obs}, the number of systems predicted by \texttt{SeBa} is 32, i.e. too small to appear in the log scale adopted in \ref{fig:Fig3}. All these systems, however, form at $Z \leq 0.02  Z_\odot$.}, while \texttt{MOBSE} predictions involve higher gas metallicity, up to $Z = 0.1 Z_\odot$. Also note that the percentage of binaries with $Z \leq 0.05 Z_\odot$ is always higher than 66\% for all the GW events. As the SFR in $6 \leq z \leq 10$ is largely dominated by galaxies with $Z \leq 0.05 \, Z_\odot$ (see Fig. \ref{fig:Fig1}), MBHBHs birth rates show the highest peak in this redshift range independently of the adopted BPS. The discrepancy in their absolute values reflect differences in the two BPS. In all metallicity bins, the number of MBHBHs predicted by \texttt{MOBSE} largely exceeds the one of \texttt{SeBa}, reflecting the assumptions made on how massive BHs form. MBHBs predicted by \texttt{MOBSE} at $Z > 0.05 Z_{\odot}$ also originate from very massive progenitors: GW150914-like systems with $Z \geq 0.08 Z_\odot$, for example, have primary stars with $m_1 > 100 M_\odot$ with sufficiently massive CO core, at the pre-SN stage, to meet the conditions of direct BH collapse, despite their mass loss \citep{Fryer2012}. Such BHs are not formed by \texttt{SeBa}, either because the IMF of the primary star does not extend beyond 100 $M_\odot$ or because efficient mass loss reduces their pre-SN mass below the $40 M_\odot$ limit, necessary for direct BH formation. Finally, it is important to stress that the histograms in Fig. \ref{fig:Fig3} result from the convolution of the intrinsic BPS metallicity distribution functions and the way metallicity-dependent formation sites evolve in the cosmological simulation. Hence, these findings indicate that both BPS models predict a fraction of MBHBs to form with large orbital separations, delaying their merger by  8 - 12 Gyr since the  formation.

\section{Conclusions}
\label{sec:Conclusion}
In this paper we investigate the origin of the most massive black hole binaries ($m_1 \ge 31 \, M_\odot$) detected during the LIGO/Virgo O1 and O2 runs (see Table \ref{tab:BBH_obs}). By running the galaxy evolution model \texttt{GAMESH} coupled with \texttt{SeBa} and \texttt{MOBSE} binary population synthesis calculations, we select binaries with primary and secondary masses and coalescence redshift within the observed ranges, and establish  their cosmological birth rate and the successive redshift evolution. We find that all birth rates peak in the redshift range $6.5 \leq z \leq 10$, i.e. before the end of cosmic reionization, regardless the binary population synthesis model. 

Three conditions act in concert to provide this result: (i) a large number of star forming dwarf galaxies contribute the total SFR in the EoR; (ii) their chemical evolution leave the gas metallicity below $Z \approx 0.1$~Z$_{\odot}$; (iii) the statistics of coalescence times of MBHBs under investigation peak at very high values ($t_c > 9.5$~Gyrs) allowing them to merge  in the interval of $z_c$ inferred from the detected GW signals \citep{Belczynski2016,Mapelli2019bnp}.

Hence, we predict these massive black hole binaries to preferentially form in low-metallicity, star forming dwarfs at redshifts significantly higher than the peak of cosmic star formation that are hardly resolved in large-scale cosmological simulations and that are beyond the observational capabilities of current electromagnetic facilities.
Future gravitational wave and electromagnetic facilities will be able to improve our knowledge of these ancient systems, fully exploiting their potential as cosmic archaeology probes.

\section*{ACKNOWLEDGMENTS }
LG and RS acknowledge support from the Amaldi Research Center funded by the MIUR program "Dipartimento di
Eccellenza" (CUP:B81I18001170001).  MM  and NG acknowledge financial support by the European Research Council for the ERC Consolidator grant DEMOBLACK, under contract no. 770017.
\bibliographystyle{mn2e}
\bibliography{MMBHO2Letter, GWbinaries}{}

\label{lastpage}
\end{document}